\RequirePackage{fix-cm}
\documentclass[twocolumn]{svjour3}          
\smartqed  
\usepackage{graphicx}
\RequirePackage[colorlinks,citecolor=blue,urlcolor=blue,linkcolor=blue]{hyperref}
\RequirePackage[numbers,sort&compress]{natbib}

\usepackage{float}
\usepackage{amsmath}
\usepackage{dblfloatfix}
\usepackage{multicol}
\usepackage{color}
\usepackage{xcolor}
\usepackage{widetext}
\usepackage{amsfonts} 
\usepackage{multirow}
\usepackage{soul}

\journalname{Eur. Phys. J. C}

\begin{document}

	
\title{Astrophysical insights into magnetic Penrose process around parameterized Konoplya-Rezzolla-Zhidenko black hole}

\titlerunning{MPP KRZ BH}



\author{Tursunali Xamidov         \and
        Sanjar Shaymatov \and
        Pankaj Sheoran \and
        Bobomurat Ahmedov 
}

\authorrunning{Xamidov \textit{et al.} }

\institute{Tursunali Xamidov \at
              Institute of Fundamental and Applied Research, National Research University TIIAME, Kori Niyoziy 39, Tashkent 100000, Uzbekistan\\
              \email{xamidovtursunali@gmail.com}           
           \and
           Sanjar Shaymatov \at
              Institute for Theoretical Physics and Cosmology, Zhejiang University of Technology, Hangzhou 310023, China\\
               Institute of Fundamental and Applied Research, National Research University TIIAME, Kori Niyoziy 39, Tashkent 100000, Uzbekistan\\
              University of Tashkent for Applied Sciences, Str. Gavhar 1, Tashkent 100149, Uzbekistan\\
              Western Caspian University, Baku AZ1001, Azerbaijan
              \email{sanjar@astrin.uz}
            \and
            Pankaj Sheoran \at
            Department of Physics, School of Advanced Sciences, Vellore Institute of Technology, Tiruvalam Rd, Katpadi, Vellore, Tamil Nadu 632014, India
            \email{pankaj.sheoran@vit.ac.in}
            \and
            Bobomurat Ahmedov \at
             Institute of Fundamental and Applied Research, National Research University TIIAME, Kori Niyoziy 39, Tashkent 100000, Uzbekistan
             \email{ahmedov@astrin.uz}
}


\date{Received: 17 August 2024 / Accepted: 27 November 2024}

\maketitle

\begin{abstract}
In this study, we investigate the parameterized Ko-noplya-Rezzolla-Zhidenko (KRZ) black hole (BH) spacetime in the presence of an external asymptotically uniform magnetic field. We first examine the innermost stable circular orbit (ISCO) radii for both neutral and charged test particles, demonstrating that the deformation parameters, $\delta_1$ and $\delta_2$, reduce the ISCO values. 
Subsequently, we assess the energy efficiency of the magnetic Penrose process (MPP) for an axially symmetric parameterized BH, analyzing the effects of the deformation parameters and the magnetic field on the energy extraction process. Our findings indicate that the rotational deformation parameter $\delta_2$ is crucial for the efficiency of energy extraction from the BH. The synergy between the rotational deformation parameter and the magnetic field significantly boosts the energy extraction efficiency, with values exceeding $100\%$. Interestingly, for extremal BHs with negative $\delta_2$ values, the energy efficiency increases, in contrast to Kerr BHs where the MPP effect diminishes. Additionally, we explore the astrophysical implications of the MPP by deriving the maximum energy of a proton escaping from the KRZ parameterized BH due to the beta decay of a free neutron near the horizon. Our results show that negative $\delta_2$ values require stronger magnetic fields to achieve equivalent energy levels for high-energy protons, providing deeper insights into high-energy astrophysical phenomena around the parameterized BH.
\end{abstract}


\section{Introduction}	
The discovery of black hole (BH) shadows and gravitational waves has marked a significant milestone in our understanding of the universe. The Event Horizon Telescope's (EHT) capture of the first direct image of a BH's shadow \cite{EventHorizonTelescope:2019dse}, along with the groundbreaking detection of gravitational waves from merging BHs by the LIGO and Virgo collaborations \cite{LIGOScientific:2016aoc,LIGOScientific:2017bnn}, has opened new frontiers in astrophysics. These achievements have not only confirmed theoretical predictions but also provided unprecedented insights into the dynamics of BHs.

For instance, the visualization of the BH shadow has allowed researchers to test general relativity in extreme conditions and study the behavior of matter in the vicinity of event horizons \cite{EventHorizonTelescope:2020qrl}. Meanwhile, the detection of gravitational waves has unveiled the violent processes of BH mergers, providing a new way to observe the universe \cite{LIGOScientific:2017ycc,LIGOScientific:2017vwq}. 
Building on these discoveries, the study of high-energy processes around BHs has gained renewed interest, particularly focusing on mechanisms like the Magnetic Penrose Process (MPP), which has the potential to explain some of the universe's most energetic phenomena.

The MPP, an extension of the original Penrose process, leverages the unique interaction between the BH's rotation and an external magnetic field. In this process, charged particles can achieve relativistic speeds due to magnetic reconnection events within the ergosphere, forming powerful jets and producing high-energy emissions. This makes the MPP particularly important for explaining observations of ultra-high-energy cosmic rays, relativistic jets, and gamma-ray bursts \cite{Wagh85ApJ}. Studies have shown that magnetic fields around BHs significantly amplify the efficiency of energy extraction, making the MPP a promising mechanism for understanding high-energy phenomena near BHs \cite{Shaymatov:2022eyz,Shaymatov24MPP1,Shaymatov24MPP2}.

Although other mechanisms, such as the Blandford-Znajek (BZ) process, have also been proposed for energy extraction, the MPP is especially notable for its ability to generate high-energy outputs through direct particle acceleration rather than relying on large-scale magnetic field structures alone. While the BZ mechanism efficiently converts a rotating BH’s energy into electromagnetic radiation and relativistic jets by twisting magnetic field lines anchored to an accretion disk, it is generally limited by the structure 
of the surrounding magnetic fields \cite{Blandford1977, Konoplya:2021qll}. In contrast, the MPP operates within the ergosphere, where charged particles are subjected to extreme gravitational and magnetic interactions, leading to highly efficient energy extraction without requiring an accretion disk. This makes the MPP potentially more versatile in environments with complex magnetic fields, such as those found in parameterized black holes like the Konoplya-Rezzolla-Zhidenko (KRZ) black holes \cite{Konoplya:2021qll}.

Future experiments, such as upgrades to the Event Horizon Telescope and the development of next-generation gravitational wave detectors like LISA (Laser Interferometer Space Antenna) and the Einstein Telescope, promise to enhance our observational capabilities \cite{Amaro:2017,2010CQGra..27s4002P}. These instruments will enable more detailed observations of BH environments, including the dynamics of accretion disks, the structure of relativistic jets, and the interaction of BHs with their surroundings. Advanced simulations and observations of BH systems will also provide critical data to understand the role of magnetic fields in high-energy astrophysical processes.

For example, projects like the Square Kilometre Array (SKA) and the Cherenkov Telescope Array (CTA) will offer new insights into the electromagnetic spectrum emissions from BHs \cite{2019arXiv191212699B,CTAConsortium:2017dvg}. These observations are crucial for testing theoretical models of energy extraction mechanisms, such as the Blandford-Znajek process and the MPP, which propose that rotating BHs can convert their rotational energy into electromagnetic radiation and relativistic jets \cite{Blandford1977,Wagh89}. By combining data from these future experiments with current observations, researchers aim to develop a more comprehensive picture of the extreme environments around BHs.

Understanding these processes not only illuminates the basic physics of BHs but also has wider consequences for high-energy astrophysics and cosmology. It might, for example, aid in the explanation of the mechanisms underlying gamma-ray bursts and other high-energy events, the generation and propagation of relativistic jets, and the origins of ultra-high-energy cosmic rays \cite{2011ARA&A..49..119K,McKinney:2012wd}. Once thought to be mysterious, isolated areas of spacetime, BHs are now essential to our comprehension of high-energy astrophysical processes.

These processes include the emission of ultra-high-energy cosmic rays, relativistic jets, and high-energy gamma-ray bursts \cite{Blandford1977,Okamoto:2005hi,Piran:2004ba}. Theoretical models and simulations have shown that BHs can act as cosmic engines, converting gravitational energy into electromagnetic radiation and kinetic energy \cite{McKinney:2008ev}. Among the various mechanisms proposed to explain these extraordinary energy outputs, the geometric Penrose process (PP) is particularly noteworthy. Initially theorized by Roger Penrose in the 1960s, this process describes how energy can be extracted from a rotating BH through the disintegration of particles within the ergosphere \cite{Penrose:1969pc}. The ergosphere, a region outside the event horizon where spacetime is dragged by the BH's rotation, allows for particle interactions that can result in one particle falling into the BH with negative energy while the other escapes with more energy than the original particle \cite{Wald:1984rg}. The gravitomagnetic monopole charge can also influence the ergosphere region, contributing to the energy extraction via the PP~\cite{Abdujabbarov11}. Later, the energy extraction process was considered for higher- dimensional BHs using the PP~\cite{Nozawa05}. Additionally, recent studies have explored the energy-extraction process in the context of quantum Schwarzschild geometry, where the process can violate the null energy condition \cite{Battista:2023iyu}. This violation opens up new possibilities for understanding the underlying physics of BHs and the mechanisms of energy extraction. Further research has also investigated superradiant energy extraction from rotating hairy Horndeski BHs \cite{Jha:2022tdl} and harvesting energy driven by the magnetic reconnection process from rotating BHs \cite{Koide_2008,Comisso21MR,Liu22ApJ,Khodadi:2022dff,Khodadi:2023juk,Shaymatov23MRP}, providing additional insights into the diverse mechanisms of energy extraction in different BH models.

The presence of a magnetic field around a BH significantly amplifies the energy extraction mechanism known as the MPP \cite{Wagh85ApJ}. This MPP leverages the interaction between the BH's rotational energy and external magnetic fields to produce extraordinary energy outputs \cite{Blandford1977,Bhat85, Komissarov:2006nz,Alic12ApJ,Moesta12ApJ}. The reconnection and twisting of magnetic fields accelerate particles to relativistic speeds, forming powerful jets that can stretch thousands of light-years from the BH's poles \cite{Blandford:1982di}. These jets not only impact the interstellar medium but also play a role in the evolution of galaxies \cite{Marscher:2008aa}. By explaining ultra-high-energy cosmic rays and high-energy gamma-ray bursts, the MPP provides insights into some of the universe's most energetic phenomena \cite{2011ARA&A..49..119K, Romero:2008zj}. Additionally, recent research highlights that magnetic fields can significantly boost the efficiency of the MPP, leading to substantial energy extraction manifesting as high-energy radiation or particle emissions \cite{Shaymatov:2022eyz,Shaymatov24MPP1,Shaymatov24MPP2}. This underscores the MPP's importance in understanding high-energy astrophysical events.

Understanding the MPP requires a comprehensive analysis of the BH's spacetime geometry and the behavior of charged particles in the presence of strong gravitational and electromagnetic fields. In this study, we focus on a specific BH solution: the parameterized Konoplya-Rezzolla-Zhidenko (KRZ) BH. This solution describes a rotating parameterized KRZ BH \cite{Konoplya_2016,Ni_2016} subjected to an external magnetic field. By examining this scenario, we aim to uncover the conditions and regions where significant energy extraction can occur, thereby enhancing our understanding of high-energy astrophysical processes. 

The parameterized KRZ BH, characterized by \textit{six deformation parameters}, provides a unique environment for studying the effects of magnetization on energy extraction processes \cite{Zhang:2024rvk}. Recent investigations into the parameters of the spherically symmetric parametrized Rezzolla-Zhidenko spacetime have provided critical insights through solar system tests, the orbit of the S2 star about Sgr $A^*$, and quasiperiodic oscillations, highlighting the broader applicability and accuracy of this parameterized model in various astrophysical contexts \cite{Shaymatov:2023jfa}. In the presence of an external magnetic field, the dynamics of charged particles around such a BH become increasingly complex. These particles can experience significant accelerations and energy gains, leading to potential high-energy phenomena observable from Earth \cite{Ruffini:1975ne,1986bhmp.book.....T}. Investigating these dynamics helps elucidate magnetic fields' role in BH environments and their impact on surrounding astrophysical processes. In this paper, we consider the MPP in the parameterized KRZ BH spacetime. By shedding light on the MPP, we aim to contribute to the broader understanding of high-energy phenomena in the vicinity of the parameterized KRZ BHs and the fundamental mechanisms driving these cosmic events. Our study not only advances theoretical knowledge but also lays the groundwork for future observational and experimental efforts to explore the extreme environments around BHs.

Our investigation is structured as follows: In Sec. \ref{Sec:Magnetized}, we delve into the geometry of the parameterized KRZ BH  spacetime, providing a detailed description of its structure and properties. This section sets the foundation for understanding the interaction between the BH and the external magnetic field. Sec. \ref{Sec:Motion} examines the motion of charged particles in this environment, analyzing their trajectories and behaviors under the influence of the combined gravitational and electromagnetic forces. This analysis is crucial for identifying the regions where the MPP can effectively take place \cite{Dovciak:2003jym,PhysRevD.74.104020}. Sec. \ref{Sec:Penrose} focuses on the MPP itself, exploring the specific mechanisms through which energy is extracted from the BH. We identify the conditions under which this process is most efficient and discuss the potential energy outputs that can be achieved. This section also highlights the significance of the MPP in explaining high-energy astrophysical phenomena \cite{1986ApJ...307...38P,1991ApJ...376..214B}. In Sec.\ref{Sec:Apl}, we discuss the astrophysical applications of the MPP. Finally, Sec.\ref{Sec:conclusion} provides a summary of our findings and discusses their broader implications for the field of astrophysics. We reflect on the potential of the MPP to revolutionize our understanding of BH physics and high-energy cosmic events.

Throughout this paper, we use a spacetime metric signature $(-,+,+,+)$ and adopt geometric units where $G = c = 1$. 

\section{Parameterized Konoplya-Rezzolla-Zhidenko BH spacetime and its electromagnetic field }\label{Sec:Magnetized}

In this section, we consider an interesting 
parameterized BH metric, usually referred to as the parameterized Konoplya-Rezolla-Zhidenko (KRZ) spacetime, which in Boyer-Lindsquist coordinates is given by \cite{Konoplya_2016,Ni_2016}. 
\begin{eqnarray}\label{Eq:metric} 
ds^2 = -\frac{N^2 - W^2 \sin^2\theta}{K^2} dt^2 - 2Wr \sin^2\theta dt d\phi+  \nonumber\\ + K^2r^2 \sin^2\theta d\phi^2 + \frac{\Sigma B^2}{N^2} dr^2 + \Sigma r^2 d\theta^2\ ,
\end{eqnarray}
where  
\begin{eqnarray}
N^2 = \left( 1 - \frac{r_0}{r} \right) \left( 1 - \frac{\epsilon_0 r_0}{r} + (k_{00} - \epsilon_0) \frac{r_0^2}{r^2} +  \frac{\delta_1 r_0^3}{r^3} \right) + \nonumber \\ +\left( \frac{a_{20}r_0^3}{r^3}+  \frac{a_{21}r_0^4}{r^4} + T \right)  \cos^2 \theta,
\end{eqnarray}

\begin{eqnarray}
    B = 1 + \frac{\delta_4 r_0^2}{r^2} + \frac{\delta_5 r_0^2}{r^2} \cos^2 \theta,
\end{eqnarray}

\begin{eqnarray}
    W = \frac{1}{\Sigma} \left( \frac{\omega_{00} r_0^2}{r^2} + \frac{\delta_2 r_0^3}{r^3} + \frac{\delta_3 r_0^3}{r^3} \cos^2 \theta \right),
\end{eqnarray}

\begin{eqnarray}
    K^2 = 1 + \frac{aW}{r} + \frac{1}{\Sigma} \left( \frac{k_{00}r_0^2}{r^2} +\left(\frac{k_{20}r_0^2}{r^2} + T \right) \cos^2 \theta\right), 
\end{eqnarray}

\begin{eqnarray}
    T=\frac{k_{21}r_0^3}{r^3 \left( 1 + \frac{k_{22} \left(1 - \frac{r_0}{r}\right)}{1 + k_{23} \left(1 - \frac{r_0}{r}\right)} \right)},
\end{eqnarray}

\begin{eqnarray}
    \Sigma = 1 + \frac{a^2}{r^2} \cos^2 \theta,
\end{eqnarray}
with the dimensionless spin parameter $a =J/M^2$ and the event horizon $r_0 = 1 + \sqrt{1 - a^2} $. We also defined following parameters 
\begin{align}
\epsilon_0 &= \frac{2 - r_0}{r_0}, & a_{20} &= \frac{2a^2}{r_0^3}, & a_{21} &= -\frac{a^4}{r_0^4} + \delta_6, \nonumber\\
\omega_{00} &= \frac{2a}{r_0^2}, & k_{00} &= k_{23} = \frac{a^2}{r_0^2}, & k_{21} &= \frac{a^4}{r_0^4} - \frac{2a^2}{r_0^3} - \delta_6,\nonumber\\ 
k_{20} &= 0, & k_{22} &= -\frac{a^2}{r_0^2}.
\end{align}

It is to be emphasized that the parameterized Konoplya-Rezolla-Zhidenko (KRZ) spacetime metric can deviate from the Kerr metric. We then consider these deviations, which can be further introduced by six deformation parameters denoted as $\{\delta_i\}$ (where $i = 1, 2, ... ,6$) that characterize the deviations. From physical point of view, these deformation parameters can be interpreted as follows:
\begin{itemize}
    \item $\delta_1 \rightarrow$ corresponds to deformations of $g_{tt}$,
    \item $\delta_2, \delta_3 \rightarrow$ correspond to rotational deformations of the metric,
    \item $\delta_4, \delta_5 \rightarrow$ correspond to deformations of $g_{rr}$,
    \item $\delta_6 \rightarrow$ correspond to deformations of the event horizon.
\end{itemize}
It is evident from the KRZ metric that it reduces to the Kerr metric exactly when considering all $\delta_i \rightarrow 0$. We note that the sign of the spin parameter $a$ can be incorporated into the first two parameters $\{\delta_2, \delta_3\}$ and by redefining the time coordinate $t$, we only consider $a > 0$. Notably, among the six deformation parameters, the first two parameters $\{\delta_1, \delta_2\}$ we aim to explore are the most crucial parameters for gaining a deeper understanding of their unique aspects and nature. Note that in the equatorial plane, all parameters are reduced, leaving only $\delta_1$ and$ \delta_2$. Therefore, for simplicity and further analysis, we shall restrict motion to the equatorial plane (i.e., $\theta=\pi/2$) and only focus on these two parameter, $\delta_1$ and $ \delta_2$. Based on the reports of the x-ray observations for the supermassive BH in Ark 564, $\delta_1$ and $\delta_2$ vary in the following range (see details \cite{Nampalliwar_2020})

\begin{equation}
    -0.27 < \delta_1 < 0.28\quad \mbox{and}\quad -0.37 < \delta_2 < 0.22\, .
\end{equation}

Examining the role of magnetic fields near a BH is crucial from an astrophysical perspective. Previous studies have highlighted the significant influence of magnetic fields on the behavior of accretion disks \cite{Blandford1977,PhysRevD.17.1518}. We further consider the magnetic field surrounding the parameterized KRZ BH. For that, the BH is considered to be immersed in an external magnetic field that is supposed to be uniform at large distances and a weak test field, satisfying the following strengths of the order of $B_1\sim 10^{8}~\rm{G}$ and $B_2\sim 10^{4}~\rm{G}$ for stellar and supermassive BHs, respectively (see, e.g., in Refs.~\cite{Piotrovich10,Baczko16,Daly:APJ:2019}). The point to note is that observational analysis through the binary BH system $V404$ Cygni~\cite{Dallilar2018} provided estimates of the magnetic field strength of the order of $B\sim 33.1 \pm 0.9~\rm{G}$. Additionally, it is worth noting that, recently, the EHT collaborations provided the estimates of the average magnetic field strength, which  is of the order of $B\sim 1 - 30~\rm{G}$ with observations at 230 GHz through the polarized synchrotron radiation around the supermassive BH at the center of the M87 galaxy (see, e.g., in Refs.~\cite{MF:2021ApJ,Narayan2021ApJ}). The magnetic field plays a pivotal role in altering a charged particle's geodesics, even though it is too small \cite{Wald:1974np}, thus causing a drastic change to the charged particles' motion. The magnetic field's influence on the motion of charged particles has since been widely developed as a well-established formalism in various scenarios~\cite[see, e.g.][]{Aliev02,Frolov10,Shaymatov20egb,Tursunov16,Shaymatov22a,Shaymatov21pdu,Shaymatov21c,Hussain17,Shaymatov22c,Shaymatov23GRG}. For our purposes, the magnetic field can be considered a weak test field in the curved background spacetime, uniform, and oriented along the axis of the BH's symmetry as that of its asymptotic properties \cite{Wald:1974np,Tursunov16,Shaymatov18a}. Therefore, we can write the vector potential as follows: 
\begin{equation}
A^{\alpha}=C_1 \xi^{\alpha}_{(t)}+ C_2 \xi^{\alpha}_{(\varphi)}\, ,
\end{equation}
where $\xi^{\alpha}_{(t)}=(\partial/\partial t)^{\alpha}$ and
axial $\xi^{\alpha}_{(\varphi)}=(\partial/\partial \phi)^{\alpha}$ are killing vectors, and $C_1$ and $C_2$ are arbitrary parameters integration constants that pertain to the feature of the field. Following the nature of an asymptotically uniform magnetic field, one can set $C_1=aB$ and $C_2=B/2$.
We then focus on the four-vector potentials of the electromagnetic field to further evaluate the efficiency of energy extraction via the MPP.  However, it turns out that the electromagnetic field's four-vector potentials become a very long and complicated expression for explicit display for the general case $\theta$. Taken all together, we consider the equatorial plane (i.e., $\theta = \pi/2$) for simplicity in obtaining the four-vector potentials of the electromagnetic field. Therefore, the four-vector potentials for the parameterized KRZ BH spacetime can be obtained as follows:  
\begin{widetext}
\begin{align}
A_t &= B\Bigg(-a\cdot \frac{r^4-r^3+a^2\left(-2+r+r^2-r r_0\right)+r r_0^2\left(2+r_0(-1+\delta_1)\right)-r_0^4\delta_1 }{r^4+a^2r(2+r)+a r_0^3\delta_2}+ 
+ \frac{a r_0^6 \delta_2^2-r r_0^3 \delta_2\left(a^2(-4+r)+r^3\right)}{2r^2[r^4+a^2r(2+r)+a r_0^3\delta_2]}\Bigg)\, , \\
A_{\phi} &=\frac{1}{2}B r^2\Bigg(1+\frac{a^2}{r^2}+\frac{a}{r}\Big(\frac{2a}{r^2}+\frac{r_0^3}{r^3}\delta_2\Big)\Bigg)-a B r \Big(\frac{2a}{r^2}+\frac{r_0^3}{r^3}\delta_2\Big)\, .
\end{align}
\end{widetext}
where $r_0 = 1+\sqrt{1-a^2}$ is the event horizon of the BH. 

We further consider charged particle dynamics around the parameterized KRZ BH immersed in an external uniform magnetic field. This is what we intend to examine in the next section. 

\section{Charged particle dynamics around the parameterized Konoplya-Rezzolla-Zhidenko BH spacetime} \label{Sec:Motion} 

In this section, we consider a charged particle motion around the parameterized KRZ BH. To this end, we utilize the Hamilton formalism for charged particles, which is defined by (see, for example~\cite{Misner73})
\begin{eqnarray}
H=\frac{1}{2}g^{\mu\nu} \left(\overline{\pi}_{\mu}-qA_{\mu}\right)\left(\overline{\pi}_{\nu}-qA_{\nu}\right)\, ,
\end{eqnarray}
where $\overline{\pi}_{\mu}$ is the canonical momentum of a charged particle and $A_{\mu}$ the four-vector potential of electromagnetic field. One can then connect the charged particle's four-momentum and its canonical momentum via the following equation
\begin{eqnarray}
p^{\mu}=g^{\mu\nu}\left(\overline{\pi}_{\nu}-qA_{\nu}\right)\, . 
\end{eqnarray}
For the motion of a charged test particle, the equation of motion can be expressed using the Hamiltonian as follows:
\begin{eqnarray} 
\label{Eq:eqh1}
  \frac{dx^\alpha}{d\lambda} = \frac{\partial H}{\partial \overline{\pi}_\alpha}\,   \mbox{~~and~~}
  \frac{d\overline{\pi}_\alpha}{d\lambda} = - \frac{\partial H}{\partial x^\alpha}\, , 
\end{eqnarray}
where $(\lambda = \tau / m)$ represents the affine parameter associated with $(\tau)$ describing the proper time for the timelike geodesics. By applying Eq.~(\ref{Eq:eqh1}), the constants of motion for timelike geodesics can be determined as follows:
\begin{eqnarray}
\label{Eq:en} \overline{\pi}_t-qA_{t}&=&
g_{tt}p^{t} + g_{t\phi}p^{\phi}\, ,\\
 \label{Eq:ln}
\overline{\pi}_{\phi}-qA_{\phi}&=& g_{\phi t}p^{t} +
g_{\phi\phi}p^{\phi}\, .
\end{eqnarray}
Note that in the equations above, $\overline{\pi}_t = -E$ and $\overline{\pi}_\phi = L$, where $E$ and $L$ represent the energy and angular momentum of a charged test particle, respectively.

\begin{figure*}[ht]
\begin{minipage}{.5\linewidth}
\centering
\includegraphics[scale=.65]{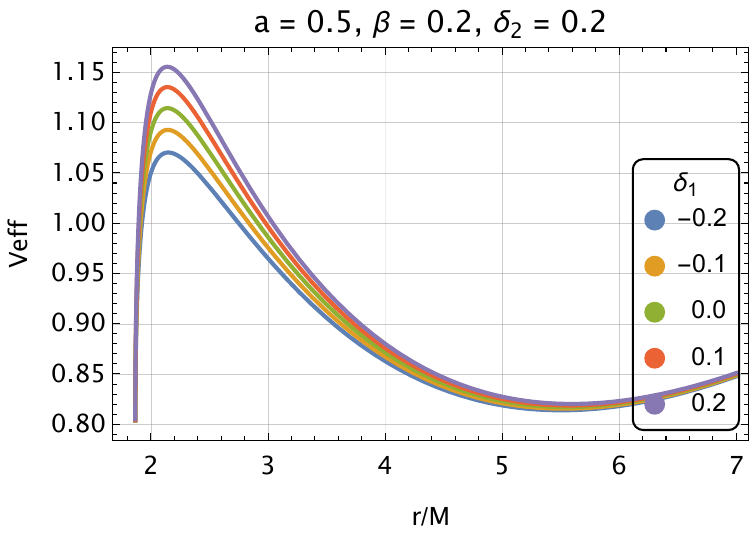}
\end{minipage}%
\begin{minipage}{.5\linewidth}
\centering
\includegraphics[scale=.65]{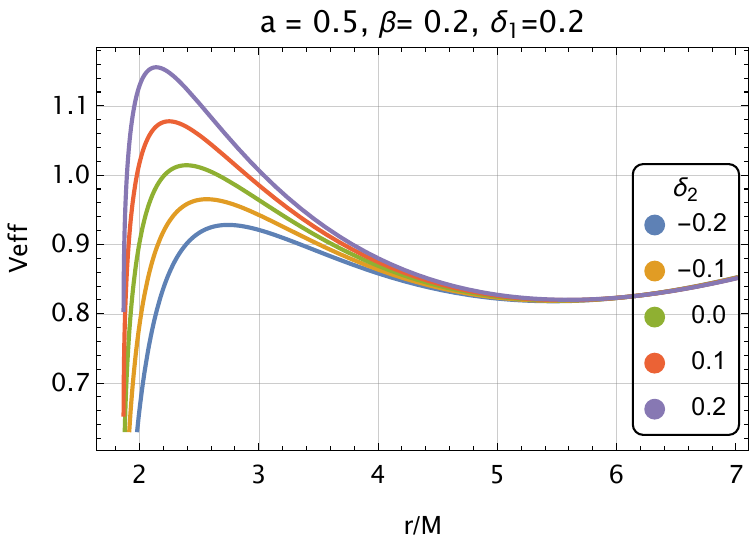}
\end{minipage}\par\medskip
\centering
\includegraphics[scale=.65]{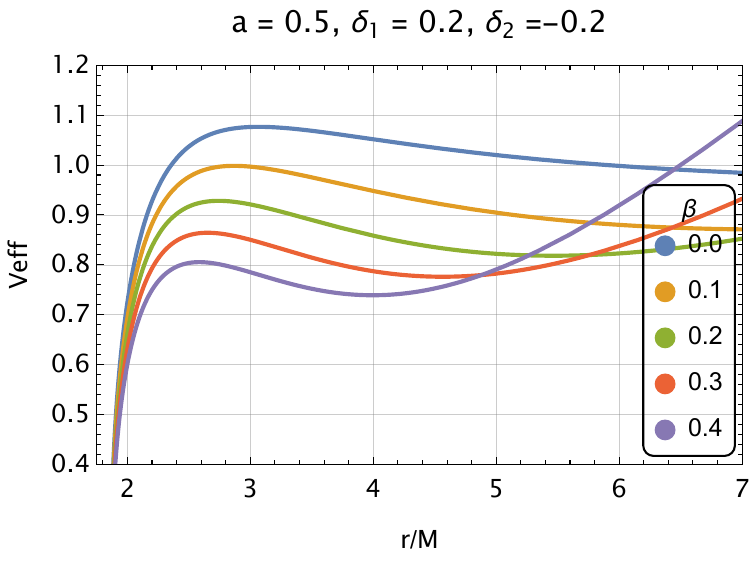}
\caption{\label{fig:effpot} 
 The radial profile of the effective potential $V_{\text{eff}}$ for different values of the deformation parameters $\delta_{1}$ (Top left panel), $\delta_{2}$ (Top right panel), and the magnetic field parameter $\beta = qB/m$ (Bottom panel). Note that we shall for simplicity restrict motion to the equatorial plane, i.e., $\theta=\pi/2$. In all cases, we have set $\delta_3 = \delta_4 = \delta_5 = \delta_6 = 0$.}
\end{figure*}

Using Eqs.~(\ref{Eq:eqh1}-\ref{Eq:ln}) and applying the normalization condition $g_{\mu\nu}p^{\mu}p^{\nu}=-m^2$, the following equation for the timelike radial motion of the charged particle in the equatorial plane (i.e., $\theta=\pi/2$) can be defined by the effective potential
\begin{align}\label{Eq:Veff}
V_{\text{eff}} &= -\frac{q}{m}A_t - \frac{g_{t\phi}}{g_{\phi\phi}} \left( \mathcal{L} - \frac{q}{m} A_\phi \right)+ 
\nonumber\\
& + \left[ \left( -g_{tt} - \omega g_{t\phi} \right) \left( 1+\frac{\left( \mathcal{L} - \frac{q}{m} A_\phi \right)^2}{g_{\phi\phi}} \right) \right]^{1/2}\, ,
\end{align}
with the angular momentum per unit mass of particle $\mathcal{L} = L/m$ and the angular frame dragging velocity $\omega = -g_{t\phi}/g_{\phi\phi}$ for a particle with zero angular momentum. It is worth noting that the effective potential plays an important role acting as a powerful tool for studying a particle motion around a BH. Based on Eq.~(\ref{Eq:Veff}), we further analyze the effective potential for nongeodesic motion of charged test particles around the parameterized KRZ BH in the presence of an external uniform magnetic field. In Fig.~\ref{fig:effpot}, we show the radial profile of the effective potential for a charged particle orbiting the parameterized KRZ BH for various possible cases of the given magnetic field parameter $\beta$ and deformation parameters $\delta_1$ and $\delta_2$. In Fig.~\ref{fig:effpot}, the top left and right panels demonstrate the impact of the deformation parameters, $\delta_{1,2}$, on the radial profile of $V_{eff}$ the effective potential, while the bottom panel  depicts the impact of the magnetic field parameter for keeping $\delta_{1,2}$ fixed. As can be seen from the top row of Fig.~\ref{fig:effpot}, there exist two extreme points, the minimum $r_{min}$ and the maximum $r_{max}$, in the effective potential, which correspond to stable and unstable circular orbits for particles around the parameterized KRZ BH. It is evident from Fig.~\ref{fig:effpot} that the shape of the effective potential is shifted upward towards higher values, but the stable orbits are shifted slightly to the left to smaller $r$ as the deformation parameters $\delta_{1,2}$ increase. Similarly, there is the same behavior for unstable orbits.  Unlike the impact of $\delta_{1,2}$, the shape of the effective potential or the maximum is shifted downward towards smaller values, causing the potential barrier to reduce. Stable orbits are significantly shifted to the right to towards the central object as a consequence of the combined effects of the magnetic field parameter and deformation parameters. This results in charged particles not escaping from the pull of gravity but instead being trapped under the gravitational and electromagnetic influences in the close vicinity of the BH.

We now examine the specific energy and angular momentum of the charged particles orbiting on the stable circular orbits, i.e., $\mathcal{E}$ and $\mathcal{L}$. To determine these quantities one needs to solve $V_{eff}={\partial V_{eff}}/{\partial r}= 0$ simultaneously. Additionally, we further consider the innermost stable circular orbits (ISCOs) for which the following standard condition 
\begin{eqnarray}
\frac{\partial^2V_{eff}}{\partial r^2}\geq 0\, ,
\end{eqnarray}
must be satisfied to determine the radius of circular orbits. We now provide a more appropriate and detailed analysis for the aforementioned parameters required for charged particles to be on the stable circular orbits around the parameterized KRZ BH. All results are tabulated in Tables~\ref{tab:table1} and \ref{tab:table2}. It can be observed from Table~\ref{tab:table1} that the ISCO radius correspondingly decreases as both the deformation parameters, $\delta_{1,2}$, increase. Similarly, $\mathcal{E}$ and $\mathcal{L}$ of the test particles orbiting on the ISCO decrease due to the influence of both parameters $\delta_{1,2}$. This suggests that the particles can lose their energy and angular momentum due to these deformation parameters when orbiting the ISCO radius around the parameterized KRZ BH. The similar behavior is also observed for these the ISCO parameters when the impact of the magnetic field parameter $\beta$ is included.             


\begin{figure}
\includegraphics[scale=0.5]{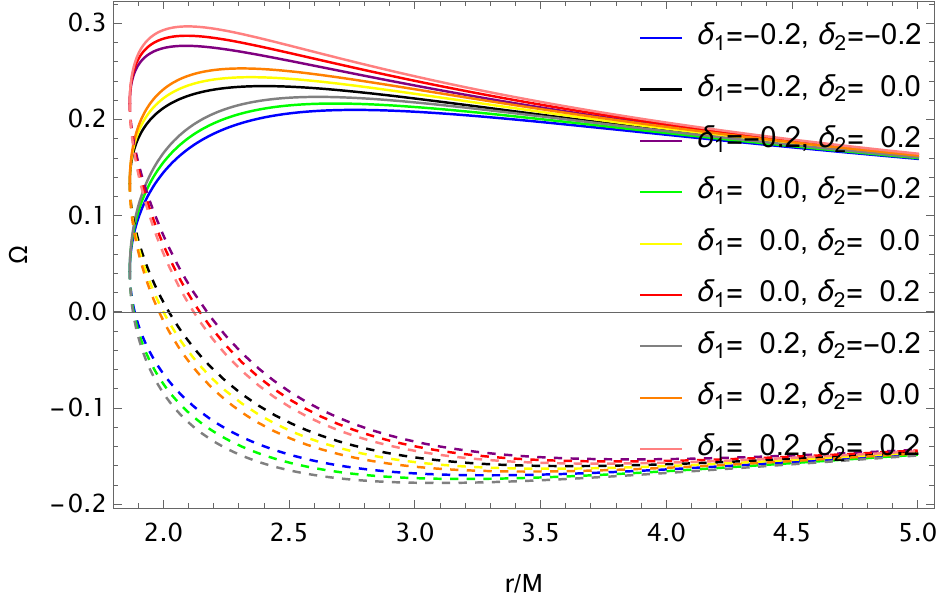}
	\caption{\label{fig:angular_vel} 
The radial behaviour of the angular velocity components $\Omega_+$ (solid lines) and $\Omega_-$ (dashed lines) for different values of the deformation parameters $\delta_{1,2}$. Note that we have used $a/M=0.5$ and $\delta_3 = \delta_4 = \delta_5 = \delta_6=0$ in this case.}
\end{figure}

We further consider a particular case for the timelike particle due to a potential barrier on the equatorial plane (i.e., $\theta=\pi/2$). To this end, we assume that the motion of the timelike particle occurs at the circular orbit satisfying $r = \text{const}$ and $\theta = \text{const})$. In this case, the four-velocity can then be defined by ${\bf u}\sim {\bf \xi}_{(t)}+\Omega {\bf \xi}_{(\phi)}$ with the angular velocity $\Omega=d\phi/dt=u^{\phi}/u^{t}$ of the particle, which can be measured by a distance observer at infinity. The angular velocity for timelike particle is then taken to be bounded as $\Omega_{-}<\Omega<\Omega_{+}$ (see details, e.g., in Refs.~\cite{1986ApJ...307...38P,Wagh:1985vuj}). Taken together, the angular velocity can be written as follows:
\begin{align}
  \Omega_{\pm} =\frac{-g_{t\phi}\pm \sqrt{(g_{t\phi})^{2}-g_{tt}g_{\phi\phi}}}{g_{\phi\phi}}\, .
\end{align}
It is to be emphasized that $\Omega_{+}$ and $\Omega_{-}$ correspond to the outer and inner photon orbit's angular velocities, respectively.  $\Omega_{+}$ is always positive, while $\Omega_{-}$ can be both positive and negative, depending on various combinations of the deformation parameters $\delta_{1,2}$, as shown in Fig.~\ref{fig:angular_vel}. In Fig.~\ref{fig:angular_vel}, we demonstrate $\Omega_{+}$ and $\Omega_{-}$ by solid and dashed lines, respectively, for various possible combinations of deformations parameters $\delta_{1,2}$. In this case of the timelike circular orbit,  the four-momentum takes the form as 
\begin{eqnarray}\label{Eq:4-mom}
\pi_{\pm}=p^{t}(1,0,0,\Omega_{\pm})\, .
\end{eqnarray}
Taking the normalization condition, $g_{\mu\nu}p^{\mu}p^{\nu}=-m^2$, into consideration together with Eq.~(\ref{Eq:4-mom}), the equation required for the circular motion of the timelike particle yields 
\begin{eqnarray}\label{Eq:W0}
\left(g_{\phi\phi}\pi_t^2+g_{t\phi}^2\right)\Omega^2&+&2g_{t\phi}\left(\pi_t^2+g_{tt}\right)\Omega\nonumber\\&+&g_{tt}\left(\pi_t^2+g_{tt}\right)=0\, ,
\end{eqnarray}
where we have denoted $\pi_t=-\left(\mathcal{E}+qA_{t}/m\right)$. 
The above equation then solves to give the angular velocity of the infalling timelike particle as (see details, for example, ~\cite{1986ApJ...307...38P,Shaymatov:2022eyz})
\begin{eqnarray}\label{29}
\Omega=\frac{-g_{t\phi}\left(\pi_t^2+g_{tt}\right)+\sqrt{\left(\pi_t^2+g_{tt}\right)\left(g_{t\phi}^2-g_{tt}g_{\phi\phi}\right)\pi_t^2}}{g_{\phi\phi}\pi_t^2+g_{t\phi}^2}\, .\nonumber\\
\end{eqnarray}

\begin{table}[]
\caption{The table shows the numerical values of the  ISCO parameters $\mathcal{L}_{ISCO}$, $\mathcal{E}_{ISCO}$, and $r_{ISCO}$ of the test particles. They are moving around KRZ parametrized plack hole on the ISCO radius. the magnetic parameter $\beta = 0$. The data is calculated for $a=0.8$ and $\theta = \pi/2$ case. }\label{tab:table1}
\scriptsize
\begin{tabular}{|c|cc|l|l|l|l|l|}
\hline
$\beta$                     & \multicolumn{2}{c|}{$\delta_1$ \textbackslash $\delta_2$}      & \multicolumn{1}{c|}{-0.2} & \multicolumn{1}{c|}{-0.1} & \multicolumn{1}{c|}{0.0} & \multicolumn{1}{c|}{0.1} & \multicolumn{1}{c|}{0.2} \\ \hline
\multirow{15}{*}{0.0} & \multicolumn{1}{c|}{\multirow{3}{*}{-0.2}} & $r_{ISCO}$ & 3.9135                   & 3.5765                   & 3.1859                  & 2.7584                 & 2.4120                  \\ \cline{3-8} 
                      & \multicolumn{1}{c|}{}                      & $\mathcal{L}_{ISCO}$ & 2.7563                   & 2.6629                   & 2.5468                   & 2.3963                  & 2.2085                  \\ \cline{3-8} 
                      & \multicolumn{1}{c|}{}                      & $\mathcal{E}_{ISCO}$ & 0.9109                  & 0.9033                    & 0.8924                 & 0.8756                 & 0.8508                 \\ \cline{2-8} 
                      & \multicolumn{1}{c|}{\multirow{3}{*}{-0.1}} & $r_{ISCO}$ & 3.7571                   & 3.4119                    & 3.0293                  & 2.6597                  & 2.3977                 \\ \cline{3-8} 
                      & \multicolumn{1}{c|}{}                      & $\mathcal{L}_{ISCO}$ & 2.6956                   & 2.5928                   & 2.4647                  & 2.3041                  & 2.1203                 \\ \cline{3-8} 
                      & \multicolumn{1}{c|}{}                      & $\mathcal{E}_{ISCO}$ & 0.9069                  & 0.8980                  & 0.8853                 & 0.8666                  & 0.8422                 \\ \cline{2-8} 
                      & \multicolumn{1}{c|}{\multirow{3}{*}{0.0}}  & $r_{ISCO}$ & 3.6054                   & 3.2617                   & 2.9066                  & 2.6017                  & 2.3969                  \\ \cline{3-8} 
                      & \multicolumn{1}{c|}{}                      & $\mathcal{L}_{ISCO}$ & 2.6312                    & 2.5187                   & 2.3804                  & 2.2156                  & 2.0402                 \\ \cline{3-8} 
                      & \multicolumn{1}{c|}{}                      & $\mathcal{E}_{ISCO}$ & 0.9024                  & 0.8922                  & 0.8778                 & 0.8581                 & 0.8351                  \\ \cline{2-8} 
                      & \multicolumn{1}{c|}{\multirow{3}{*}{0.1}}  & $r_{ISCO}$ & 3.4634                   & 3.1329                   & 2.8181                  & 2.5691                  & 2.4031                  \\ \cline{3-8} 
                      & \multicolumn{1}{c|}{}                      & $\mathcal{L}_{ISCO}$ & 2.5631                   & 2.4417                   & 2.2962                  & 2.1325                  & 1.9670                  \\ \cline{3-8} 
                      & \multicolumn{1}{c|}{}                      & $\mathcal{E}_{ISCO}$ & 0.8975                  & 0.8860                   & 0.8703                 & 0.8506                 & 0.8292                 \\ \cline{2-8} 
                      & \multicolumn{1}{c|}{\multirow{3}{*}{0.2}}  & $r_{ISCO}$ & 3.3356                   & 3.0289                  & 2.7573                  & 2.5519                  & 2.4131                  \\ \cline{3-8} 
                      & \multicolumn{1}{c|}{}                      & $\mathcal{L}_{ISCO}$ & 2.4921                   & 2.3631                   & 2.2143                  & 2.0550                  & 1.8997                  \\ \cline{3-8} 
                      & \multicolumn{1}{c|}{}                      & $\mathcal{E}_{ISCO}$ & 0.8922                  & 0.8796                  & 0.8632                & 0.8440                 & 0.8244                 \\ \hline
\end{tabular}
\end{table}

\begin{table}[]
\caption{The table shows the numerical values of the  ISCO parameters $\mathcal{L}_{ISCO}$, $\mathcal{E}_{ISCO}$, and $r_{ISCO}$ of the test particles. They are moving around KRZ parametrized plack hole on the ISCO radius. the magnetic parameter $\beta = 0.01$. The data is calculated for $a=0.8$ and $\theta = \pi/2$ case. }\label{tab:table2}
\scriptsize
\begin{tabular}{|c|cc|l|l|l|l|l|}
\hline
$\beta$                     & \multicolumn{2}{c|}{$\delta_1$ \textbackslash $\delta_2$}      & \multicolumn{1}{c|}{-0.2} & \multicolumn{1}{c|}{-0.1} & \multicolumn{1}{c|}{0.0} & \multicolumn{1}{c|}{0.1} & \multicolumn{1}{c|}{0.2} \\ \hline
\multirow{15}{*}{0.01} & \multicolumn{1}{c|}{\multirow{3}{*}{-0.2}} & $r_{ISCO}$ & 3.8995                    & 3.5656                   & 3.1777                  & 2.7525                   & 2.4075                  \\ \cline{3-8} 
                      & \multicolumn{1}{c|}{}                      & $\mathcal{L}_{ISCO}$ & 2.7249                   & 2.6339                   & 2.5206                   & 2.3733                  & 2.1888                  \\ \cline{3-8} 
                      & \multicolumn{1}{c|}{}                      & $\mathcal{E}_{ISCO}$ & 0.8978                  & 0.8906                  & 0.8802                 & 0.8640                 & 0.8400                 \\ \cline{2-8} 
                      & \multicolumn{1}{c|}{\multirow{3}{*}{-0.1}} & $r_{ISCO}$ & 3.7446                   & 3.4024                   & 3.0224                  & 2.6547                   & 2.3935                  \\ \cline{3-8} 
                      & \multicolumn{1}{c|}{}                      & $\mathcal{L}_{ISCO}$ & 2.6654                   & 2.5652                   & 2.4401                  & 2.2827                  & 2.1022                  \\ \cline{3-8} 
                      & \multicolumn{1}{c|}{}                      & $\mathcal{E}_{ISCO}$ & 0.8941                  & 0.8857                  & 0.8735                 & 0.8554                 & 0.8317                 \\ \cline{2-8} 
                      & \multicolumn{1}{c|}{\multirow{3}{*}{0.0}}  & $r_{ISCO}$ & 3.5946                   & 3.2537                    & 2.9008                  & 2.5972                  & 2.3929                  \\ \cline{3-8} 
                      & \multicolumn{1}{c|}{}                      & $\mathcal{L}_{ISCO}$ & 2.6021                   & 2.4925                   & 2.3573                  & 2.1959                  & 2.0235                  \\ \cline{3-8} 
                      & \multicolumn{1}{c|}{}                      & $\mathcal{E}_{ISCO}$ & 0.8899                  & 0.8802                  & 0.8664                 & 0.8473                 & 0.8249                 \\ \cline{2-8} 
                      & \multicolumn{1}{c|}{\multirow{3}{*}{0.1}}  & $r_{ISCO}$ & 3.4540                     & 3.1261                   & 2.8131                  & 2.5650                  & 2.3992                  \\ \cline{3-8} 
                      & \multicolumn{1}{c|}{}                      & $\mathcal{L}_{ISCO}$ & 2.5354                   & 2.4169                   & 2.2747                  & 2.1142                  & 1.9515                  \\ \cline{3-8} 
                      & \multicolumn{1}{c|}{}                      & $\mathcal{E}_{ISCO}$ & 0.8853                  & 0.8743                  & 0.8592                 & 0.8402                 & 0.8194                 \\ \cline{2-8} 
                      & \multicolumn{1}{c|}{\multirow{3}{*}{0.2}}  & $r_{ISCO}$ & 3.3275                   & 3.0230                   & 2.7528                  & 2.5481                   & 2.4092                  \\ \cline{3-8} 
                      & \multicolumn{1}{c|}{}                      & $\mathcal{L}_{ISCO}$ & 2.4658                   & 2.3399                   & 2.1943                  & 2.0381                   & 1.8853                  \\ \cline{3-8} 
                      & \multicolumn{1}{c|}{}                      & $\mathcal{E}_{ISCO}$ & 0.8804                  & 0.8682                  & 0.8525                 & 0.8339                  & 0.8148                  \\ \hline
\end{tabular}
\end{table}

\section{Energy extraction through the magnetic Penrose process}\label{Sec:Penrose}

As mentioned earlier, astronomical observations have revealed high-energy cosmic rays coming from active galactic nuclei(AGN)~\cite{Fender04mnrs,Auchettl17ApJ,IceCube17b}. 
In order to propose a model that explains the results of these observations, it is valuable to study the geometric PP. This process was first theoretically described by Penrose in 1969 \cite{Penrose:1969pc}. Since then, different models have been developed to explain the energy extraction mechanisms from BHs and jets in AGNs~\cite{McKinney07}. The MPP \cite{Blandford1977} has important consequences for high-energy astrophysical events, generalizing the geometric PP and including the generation of relativistic jets observed in AGN and gamma-ray bursts (GRBs). Due to the presence of the ergoregion located outside the event horizon but inside the static radius, energy can be extracted from the rotating BH in the PP. In this scenario, a massive particle entering the ergosphere is divided into two fragments. The momentum of these fragments allows one to escape to infinity, while the other is pulled into the event horizon. After this process, the escaping particle can have more energy than the initial particle. Therefore, it has been well considered energy extraction from the BH.

Let us then turn to this thought mechanism by supposing a neutral particle, $q_1=0$, that falls into a BH and splits into two particles in the ergoregion. The incident particle with $m_1$ has energy $E_1\geq 1$, while the other two particles formed after splitting have energy and charge, such as $\left(E_2, q_2\right)$ and $\left(E_3, q_3\right)$. We consider that the Second particle with the mass $m_2$ falls into BH with energy $E_2<0$, while the third particle with the mass $m_3$ escapes from the BH with energy $E_3=E_1-E_2>E_1$. In this scenario, we can write the following conservation laws for the process happening in the ergosphere of the BH  \begin{eqnarray}\label{Eq:con_laws}
E_1=E_{2}+E_{3}\,
\mbox{~~~and~~~}
L_1=L_{2}+L_{3}\, ,    
\end{eqnarray}
together with $m_1=m_{2}+m_{3}$ and $q_{2}=-q=-q_{3}$. Taken together, the four-momentum can be written as follows \cite{Blandford1977,Shaymatov:2022eyz}:
\begin{eqnarray}\label{Eq:con_law}
m_1u_1^{\mu}&=& m_2u_2^{\mu}+m_3u_3^{\mu}\, .
\end{eqnarray}
including the conservation law for all the components of four-momentum. Considering $u_i^{\phi}=\Omega\, u_i^{t}=-\Omega \mathcal{A}_i/\mathcal{B}_i$ we recall Eq.~(\ref{Eq:con_law}), which yields
\begin{eqnarray}
\Omega_1m_1\mathcal{A}_{1}\mathcal{B}_2\mathcal{B}_3=\Omega_2m_2\mathcal{A}_{2}\mathcal{B}_3\mathcal{B}_1+\Omega_3m_3\mathcal{A}_{3}\mathcal{B}_2\mathcal{B}_1\, ,
\end{eqnarray}
where  $\mathcal{A}_i=\mathcal{E}_i +q_iA_{t}/m_i$ and $\mathcal{B}_i=g_{tt}+g_{t\phi} \Omega_i $. After making some algebraic manipulation, we obtain the following equation
\begin{eqnarray}
\frac{E_3+q_3A_{t}}{E_1+q_1A_{t}}=\left(\frac{\Omega_1\mathcal{B}_2-\Omega_2\mathcal{B}_1}{\Omega_3\mathcal{B}_2-\Omega_2\mathcal{B}_3}\right)\frac{\mathcal{B}_3}{\mathcal{B}_1}\, ,
\end{eqnarray}
allowing the energy of escaping particle to be written as 
\begin{eqnarray}\label{Eq:E3E1}
E_3=\chi\left(E_1+q_1A_{t}\right)-q_3A_{t}\, .
\end{eqnarray}
To have simpler form of Eq.~(\ref{Eq:E3E1}), we have defined the following notations
\begin{eqnarray}\label{Eq:chi}
\chi=\left(\frac{\Omega_1-\Omega_2}{\Omega_3-\Omega_2}\right)\frac{\mathcal{B}_3}{\mathcal{B}_1}
\end{eqnarray}
with 
\begin{eqnarray}
\Omega_1= \Omega\, , \mbox{~~} \Omega_2=\Omega_{-}\, \mbox{~~and~~} \Omega_3=\Omega_{+}\, .
\end{eqnarray}
\begin{figure*}
\includegraphics[scale=0.65]{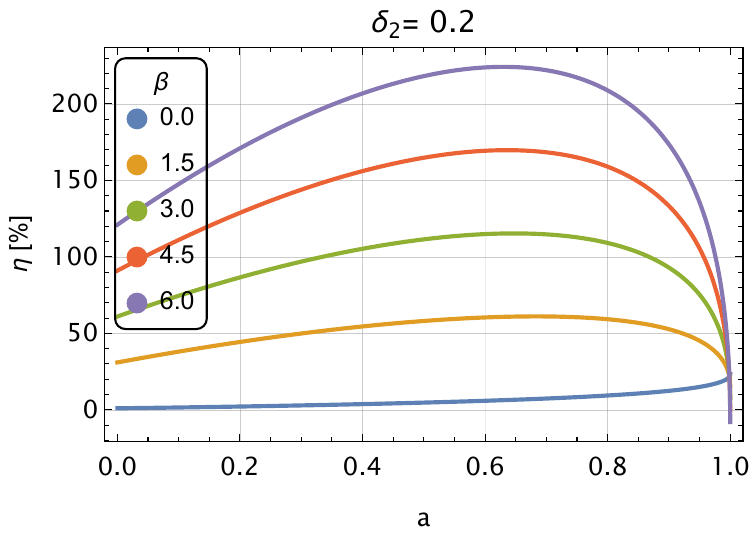}
\includegraphics[scale=0.65]{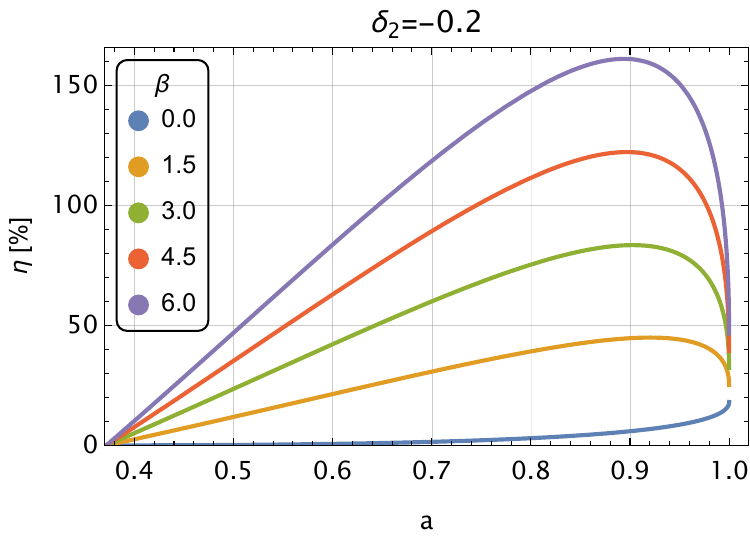}
\includegraphics[scale=0.65]{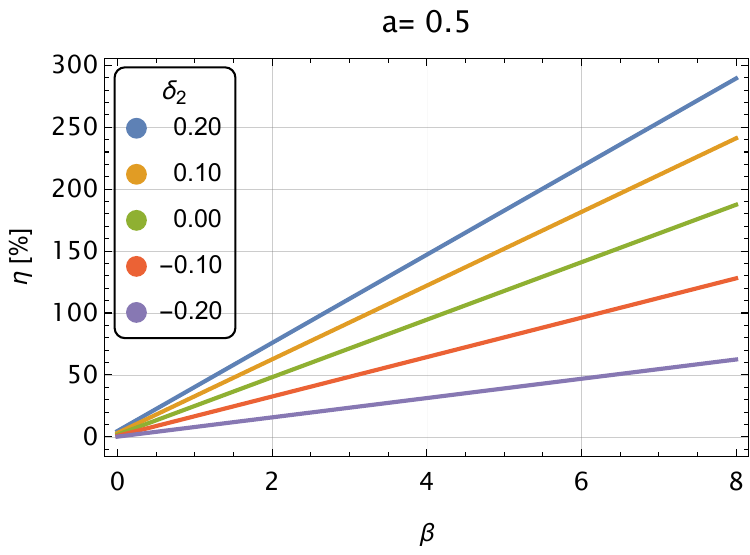}
\includegraphics[scale=0.65]{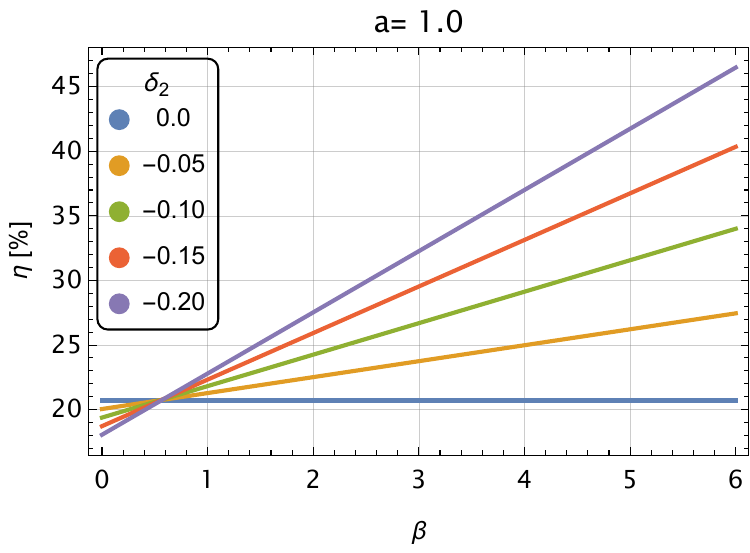}
\caption{\label{fig:en_eff1} The efficiency of energy extraction from the parameterized KRZ BH using the MPP. Top row: $\eta$ is plotted as a function of the rotation parameter $a$ for positive $\delta_2=0.2$ (left) and negative $\delta_2=-0.2$ (right) in the equatorial plane (i.e., $\theta=\pi/2$) of the BH. Bottom row: $\eta$ is plotted as a function of the magnetic field parameter $\beta$ for various possible combinations of $\delta_2$ for fixed $a=0.5$ (left) and the extremal value, $a=1$ (right). }
\end{figure*}

We then turn to the efficiency of energy extraction from the BH through the MPP, which can generally be defined as follows:
\begin{eqnarray} \label{Eq:eta}
\eta=\frac{E_3-E_1}{E_1}\, .
\end{eqnarray}
Taken Eq.~(\ref{Eq:E3E1}) together,  we rewrite Eq.~(\ref{Eq:eta}) as  
\begin{eqnarray}
\eta= \chi-1-\frac{q_3A_t}{m_1\pi_{t1}+q_1A_t}+\frac{ q_1A_t}{m_1\pi_{t1}+q_1A_t}\,\chi\, .
\end{eqnarray}
which defines the energy efficiency of the escaping charged particle, $q_2$. 
Keeping our initial assumption in mind in this scenario, i.e., $q_1=0$ satisfying the conservation of charge, $q_2+q_3=0$, the efficiency of energy extraction form the parameterized KRZ BH via the MPP can be defined as follows:
\begin{eqnarray} \label{Eq:EnerEff}
\eta= \left(\frac{\Omega-\Omega_{-}}{\Omega_{+}-\Omega_{-}}\right)\left(\frac{g_{tt}+\Omega_{+}g_{t\phi}}{g_{tt}+\Omega\,g_{t\phi}}\right)-1-\frac{q_3A_t}{E_1}\, .
\end{eqnarray}
In this scenario, the efficiency of energy extraction approaches its maximum, if and only if the splitting process occurs very near the BH's event horizon. Therefore, Eq.~(\ref{Eq:EnerEff}) for the energy efficiency of the escaping particle takes following form: 
\begin{align} \label{Eq:EneffGeneral}
\eta &= \frac{1}{2} \left(\sqrt{\frac{(2 a+ r_0^2\delta_2)^2}{r_0^4+a^2 r_0(r_0+2)+a  r_0^3 \delta_2}+1}-1\right) + 
\nonumber\\
& + \beta\frac{(2 a+r_0^2 \delta_2 ) [r_0^3+a^2 (r_0-2)-a r_0^2\delta_2 ]}{2 (r_0^4+a^2 r_0(r_0+2)+a r_0^3\delta_2)}
\end{align}
where we have defined $\beta = q_3 B/E_1\sim q B/m$ as the magnetic parameter highlighting the impact of the MPP on the efficiency of energy extraction and $r_0=1+\sqrt{1-a^2}$ as the horizon radius of the parameterized KRZ BH.

We now turn to analyze the efficiency of energy extraction from the parameterized KRZ BH.  In Fig.~\ref{fig:en_eff1}, we show the efficiency of energy extraction from the BH as a function of the spin parameter $a$ in the top row, and the magnetic field parameter $\beta$ in the bottom row. It is observed from the top row of Fig.~\ref{fig:en_eff1} that the shape of the efficiency of energy extraction shifts upward toward larger values as the magnetic field parameter $\beta$ increases. Therefore, the efficiency is strongly enhanced due to the influence of the magnetic field parameter, resulting in it exceeding $100\%$. This enhancement occurs due to the MPP and allows for arbitrarily large energy efficiency.  Additionally, the point to note is that the efficiency of energy extraction with positive $\delta_2>0$ takes larger values than with negative $\delta_2<0$. Notably, it can be observed from the top left panel of Fig.~\ref{fig:en_eff1} that the energy efficiency reaches $\eta\sim23.3\%$, which is greater than the Kerr case (where it is $\eta\sim20.7\%$) when $a\to a_{ext}$. This occurs because the MPP part goes to $\eta\vert_{q\neq0}= 0$ when $a\to a_{ext}$. However, the negative values of the deformation parameter $\delta_2$ allow the MPP part to retain its contribution even when $a\to a_{ext}$, as shown in the right column of Fig.~\ref{fig:en_eff1}. This is a remarkable and distinguishing nature of these positive and negative values of the deformation parameter $\delta_2$. In the bottom row of Fig.~\ref{fig:en_eff1}, we demonstrate the impact of the deformation parameter $\delta_2$ on the efficiency of energy extraction from the BH via the MPP. It is evident from the left panel of Fig.~\ref{fig:en_eff1} that the curves of the efficiency of energy extraction shift upward toward larger values and surpass $\eta>100\%$ as we increase the deformation parameter $\delta_2$ from negative to positive values. One can also notice that the efficiency becomes larger than the Kerr case for positive values of $\delta_2$, but less for its negative values, as depicted in the bottom left panel of Fig.~\ref{fig:en_eff1}. As highlighted earlier, the efficiency of energy extraction increases with the rise in the negative value of $\delta$ even in the case of $a\to a_{ext}$; see the bottom right panel of Fig.~\ref{fig:en_eff1}. This is one of the unique aspects of the deformation parameter $\delta_2$.

\section{Astrophysical applications of the magnetic Penrose process \label{Sec:Apl}}

In this section, we consider astrophysical applications of the MPP. To this end, we estimate the maximum energy of a proton escaping from the ergoregion of the parameterized KRZ BH. Unlike the previous analysis of the efficiency of energy extraction, we determine the energy after the proton gets accelerated by the magnetic field in the ergoregion of the parameterized KRZ BH. This results in the MPP being considered for possible applications of the results presented in the previous section. To provide a more appropriate and quantitative analysis, we need to examine the neutron beta-decay process in the close vicinity of the parameterized KRZ BH \cite{PierreAuger:2018qvk,Tursunov:2020juz,2022Symm...14..482T}:
\begin{eqnarray}
    n^0 \rightarrow p^+ + W^- \rightarrow p^+ + e^- + \overline{\nu}_e\, .
\end{eqnarray}
We now turn to determine the energy of accelerating protons. From Eqs.~(\ref{Eq:eta}) and (\ref{Eq:EneffGeneral}), we obtain the analytical form of the energy for the accelerating protons: 
\begin{figure*}
\includegraphics[scale=0.435]{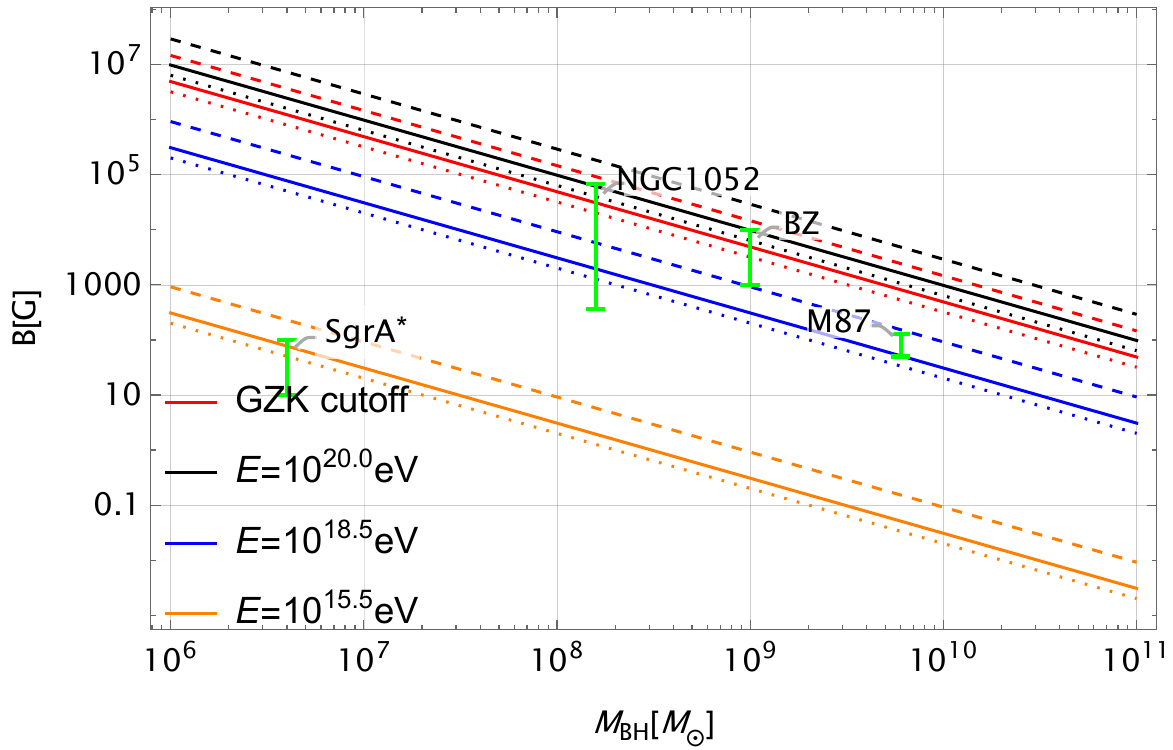}
\includegraphics[scale=0.52]{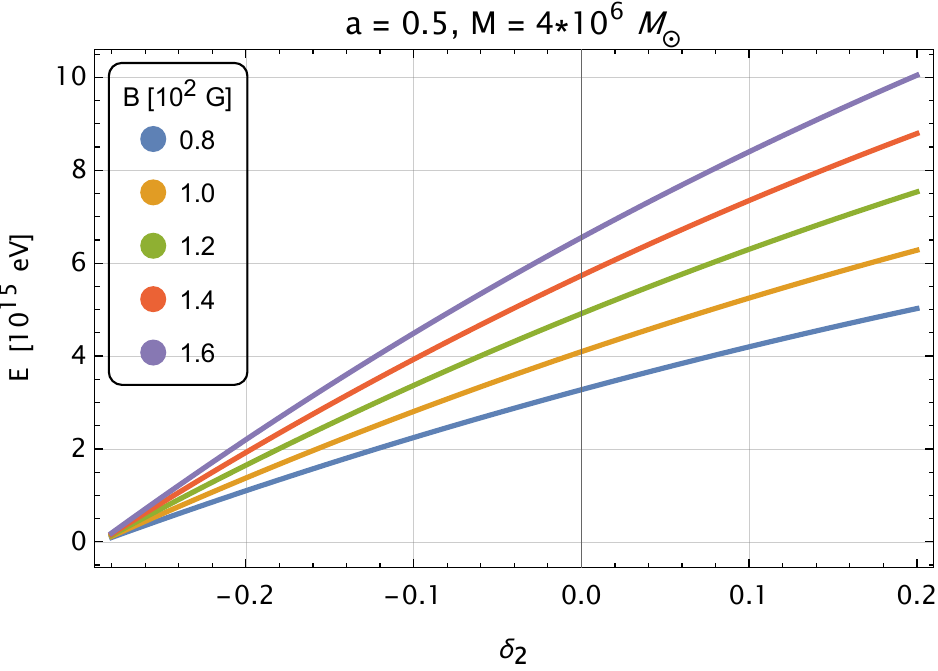}
\caption{\label{fig:en_eff} The left panel shows constraint plot of the BH mass and magnetic field for some selected BH candidates serving as sources of high-energy protons with different energies for various possible cases of the rotational deformation parameter $\delta_2$. Solid lines correspond to $\delta_2=0.0$, dotted to $\delta_2=0.2$, and dashed to $\delta_2=-0.2$. The right panel shows the energy of accelerated protons by SgrA$^{\star}$ after beta decay as a function of $\delta_2$ for various possible values of the magnetic field strength $B$. }
\end{figure*}
\begin{align} \label{Eq:E3}
E_{p^+} &= \frac{1}{2} \left(\sqrt{\frac{(2 a+ r_0^2\delta_2)^2}{r_0^4+a^2 r_0(r_0+2)+a  r_0^3 \delta_2}+1}-1\right) m_{n^0}c^2 + 
\nonumber\\
& + e B\frac{(2 a+r_0^2 \delta_2 ) [r_0^3+a^2 (r_0-2)-a r_0^2\delta_2 ]}{2 (r_0^4+a^2 r_0(r_0+2)+a r_0^3\delta_2)}\, ,
\end{align}
where $m_{n^0}$ refers to the mass of falling neutron. 
Using Eq.~(\ref{Eq:E3}), the energy of the escaping proton after the beta-decay of a free neutron can be determined by the following expression:
\begin{eqnarray}
E_{p^+} = 1.57 \times 10^{20} \text{eV} \left( \frac{B}{10^4 \text{G}} \right) \left( \frac{M}{10^9 M_\odot} \right) \left( \frac{a}{0.5} \right)\, , 
\end{eqnarray}
where we have set the rotational deformation parameter $\delta_2 = 0.2$ of the parameterized metric considered here. To be more quantitative, we estimate this energy of escaping protons. We show that it can be estimated to be $E_{p^+}$ reaches the value $10^{20} eV$ for given the mass $M \sim 10^9 M_\odot$ and magnetic field strength $B \sim 10^4 \text{G}$ provided that the beta decay occurs around the BH, especially very close to the BH's event horizon, i.e., $r_{\text{decay}} \approx r_0$. This value of energy corresponds to the ultra-high-energy cosmic rays.

We now turn to estimate the accelerating power of the supermassive BH at the center of SgrA$^{\star}$ with a mass of $4 \times 10^6 M_\odot$ (see, for example, In Ref.~\cite{Parsa:p2017}) and the magnetic field of nearly 1 to 200G at the event horizon scales (see, for example, in Refs.~\cite{Eckart_2012,Eatough_2013,MF:2021ApJ,Narayan2021ApJ}). After beta decay the maximum energy of protons accelerated by SgrA$^{\star}$ as the parameterized KRZ BH can be estimated as follows:
\begin{eqnarray}
E_{p^+}^{\text{SgrA*}} &=& 6.27 \times 10^{15} \text{eV} \left( \frac{B}{10^2 \text{G}} \right) \nonumber\\&&\times\left( \frac{M}{4 \times 10^6 M_\odot} \right) \left( \frac{a}{0.5} \right)\, ,
\end{eqnarray}
referred to as the knee of the cosmic ray energy spectrum. The knee energy is located around energies of $10^{15} \text{---} 10^{16} \, \text{eV}$. After this point, the flux of cosmic rays suddenly decreases. It is evident from the above estimated value that the knee energy for the parameterized KRZ BH is larger than the one for the Kerr BH case \cite{Tursunov:2020juz}.   

We now provide a more appropriate and detailed analysis associated with constraints on the BH mass and the strength of the magnetic field existing in the surrounding environment of BHs serving as sources of highly energetic astrophysical phenomena, such as high-energy protons with various energies. In Fig.~\ref{fig:en_eff}, we show how the combined effects of the magnetic field and BH mass contribute to the energy of escaping protons which can likely be considered ultra-high-energy cosmic rays. As can be seen from Fig.~\ref{fig:en_eff}, the vertical green lines manifest the range of possible magnetic field strength as 
stated by the observational data for BH candidates. For example, measurements of magnetic fields have been conducted with precise and various confidential methods for some selected sources (see details, e.g., in Refs.~\cite{Baczko_2016,Eckart_2012,Kino_2015,Eatough_2013}). Additionally, EHT collaborations have reported very recently that the magnetic field strength is of the order of $B\sim 1 - 30~\rm{G}$ in the emission region, according to observations at 230 GHz around the supermassive BH at the center of the M87 galaxy~\cite{MF:2021ApJ,Narayan2021ApJ}. Based on these observational data for the selected sources, we utilize the range of possible magnetic field strength in our analysis, as seen in Fig.~\ref{fig:en_eff}. We also note that solid black, blue and orange lines depict specific energies of the escaping particles in the limit of $\delta_2=0$, thus manifesting the acceleration capability of a Kerr BH. Unlike solid lines, the dashed and dotted lines depict the effect of the deformation parameter $\delta_2$ on the acceleration capability of the parameterized KRZ BH. Interestingly, we find that for the proton to get accelerated with the same energy under the acceleration capability of the parameterized KRZ BH, it is required for a BH to have more mass and a stronger magnetic field for the negative case of the deformation parameter $\delta_2<0$. However, the opposite is true for the positive values of the deformation parameter $\delta_2>0$. The point to be noted here is that we also show the red lines to describe the Greisen-Zatsepin-Kuzmin (GZK) limit, usually referred to as cutoff effect, for various possible values of the parameter $\delta_2$. The GZK cutoff limit delineates the maximum energy protons can have when travelling from other galaxies through the intergalactic medium to our SgrA$^{\star}$ galaxy. This limit can be estimated theoretically to be nearly 5$\times10^{19}\,eV$ \cite{Greisen1966}. The reason for the existence of this limit is due to interactions between the protons and the microwave background radiation over vast distances. In addition, in the right panel of Fig.~\ref{fig:en_eff}, we demonstrate the energy of accelerated protons as a function of the deformation parameter $\delta_2$ for various possible combinations of the magnetic field strength $B$. From the right panel of Fig.~\ref{fig:en_eff}, the energy of accelerated protons rises as $\delta_2$ increases for given BH parameters. It can be observed from the right panel that the curves of energy shift upward toward larger values as a consequence of the rise in the value of the magnetic field strength. One can infer from the results that the energy of accelerated protons in the ergoregion of the parameterized KRZ BH is more sensitive to the background magnetic field $B$ and the rotational deformation parameter $\delta_2$ as well.

\section{Conclusions}
\label{Sec:conclusion}

In this paper, we examined the parameterized KRZ black hole (BH) spacetime within an external asymptotically uniform magnetic field. By applying the MPP, we demonstrated how the deformation parameter, $\delta$, affect the efficiency of energy extraction from the BH. Our findings provide new insights into the MPP, enhancing our understanding of high-energy phenomena near parameterized KRZ BHs and the fundamental mechanisms driving these events. This study not only advances theoretical knowledge but also lays the groundwork for future observational and experimental research, aiming to explore the extreme environments around BHs more comprehensively.

We began by deriving the field equations and found the electromagnetic four-vector potentials for the case of the parameterized KRZ BH. We then investigated the motion of charged test particles around the BH immersed in an external uniform magnetic field. We showed that the stable orbits are shifted to the left to smaller $r$ due to the rise in the value of the deformation parameters $\delta_{1,2}$. We also showed that the deformation parameters $\delta_{1,2}$ strengthen the potential barrier. Additionally, we found that charged particles are not able to escape from the pull of gravity but instead to be trapped under the combined effects of gravitational and electromagnetic forces around the BH.

From an astrophysical point of view, astrophysical rotating BHs are considered the high-energy sources in the universe. Therefore, testing BH's energetic properties is valuable through the various methods for understanding of their rotational energy. In this paper, we also considered the MPP to study the efficiency of energy extraction from the parameterized KRZ BH. We now turn to analyze the efficiency of energy extraction from the parameterized KRZ BH.  We showed the efficiency of energy extraction is strongly enhanced by the MPP. This results in the energy efficiency to surpass $100\%$. This enhancement permits for arbitrarily large energy efficiency due to the MPP. Additionally, we observed that the efficiency of energy extraction with positive $\delta_2>0$ is larger than negative $\delta_2<0$. Interestingly, it was found that the energy efficiency approaches $\eta\sim23.3\%$, which is greater than the Kerr case, as the MPP part goes to $\eta\vert_{q\neq0}= 0$ when $a\to a_{ext}$. It is worth noting, however, that the negative values of the deformation parameter $\delta_2$ allow the MPP part to retain its contribution regardless of the extremal value $a_{ext}$ of the rotation parameter. This is a remarkable and distinguishing feature of these positive and negative values of the deformation parameter $\delta_2$. Furthermore, it is observed that the efficiency of energy extraction surpasses $\eta>100\%$ as the deformation parameter $\delta_2$ increases from its negative to positive values. Notably, we showed that the efficiency is larger than the Kerr case for $\delta_2>0$, but less for $\delta_2<0$. It must be noted that for the negative values of $\delta$ the MPP part does not go to zero even when $a\to a_{ext}$, thereby increasing the energy efficiency, i.e.., $\eta(\delta_2\neq 0)|_{a\to a_{ext}}>\eta(\delta_2=0)|_{a\to a_{ext}}$. This is one of the unique aspects of $\delta_2$ in the parameterized KRZ BH spacetime.   

We also provided a more appropriate and detailed analysis associated with constraints on the BH mass and the strength of the magnetic field around selected BHs serving as sources of high-energy protons with various energies. We showed the combined effects of the magnetic field and BH mass on the energy of escaping protons considered as ultra-high-energy cosmic rays. Based on the observational data for some selected sources, we used the range of possible magnetic field strengths in our analysis. We demonstrated deviations from the acceleration capability of a Kerr BH due to the deformation parameter $\delta_2$ of the parameterized KRZ BH. Interestingly, we found that more mass and a stronger magnetic field are required for the proton to be accelerated to the same energy under the acceleration capability of the parameterized KRZ BH for the case of a negative deformation parameter $\delta_2<0$. However, the opposite is true for the case of a positive $\delta_2>0$. We also analyzed the GZK cutoff energy limit and showed the influence of the deformation parameter  $\delta_2$. Notably, the GZK cutoff limit can also be affected in a  similar way in the parameterized KRZ BH spacetime. Furthermore, we studied how the energy of accelerated protons can be influenced under the combined effects of the deformation parameter $\delta_2$ and the magnetic field$B$. We showed that the energy of accelerated protons increases with an increase in the deformation parameter $\delta_2$ and the magnetic field strength $B$. The results suggest that the energy of accelerated protons is more sensitive to the background magnetic field $B$ and the rotational deformation parameter $\delta_2$ of the parameterized KRZ BH spacetime. 

Astrophysically, it is often difficult for distant observers to differentiate between various BH geometries due to the similar radiation emitted by accretion disks in low-mass X-ray binaries. This similarity is a result of the inherent degeneracy in BH spacetimes. Based on these theoretical studies, 
the approach we utilized here is vital for uncovering the distinct and novel characteristics of the parameterized KRZ BH spacetime, providing deeper insights into its unique properties.

\begin{acknowledgements}
This work is supported by the National Natural Science Foundation of China under Grant No. 11675143 and the National Key Research and Development Program of China under Grant No. 2020YFC2201503. PS acknowledges the Vellore Institute of Technology for providing financial support through its Seed Grant (No. SG20230079), year 2023. 
\end{acknowledgements}

\bibliographystyle{spphys}       
\bibliography{MPP_KRZ_BH}
		
\end{document}